\documentclass[aip, apl, reprint]{revtex4-1}

	\usepackage{amsmath,latexsym}
	\usepackage{graphicx}
	\usepackage{dcolumn}
	\usepackage{bm}
	
	\usepackage[utf8]{inputenc}
	\usepackage[T1]{fontenc}
	\usepackage{mathptmx}
	\usepackage{etoolbox}
	
\begin{document}
	
	\title{Time-Domain Reconstruction of the Speed of Sound in Ring-Array Ultrasound Computed Tomography with Randomized Super-Shots}
	
	\author{Luca A. Forte}
	\email{lforte001@dundee.ac.uk}
		\affiliation{School of Science and Engineering, University of Dundee, Nethergate, Dundee, DD1 4HN, Scotland, UK}
	
	\date{\today}
	
	\begin{abstract}
	Ring-array ultrasound computed tomography has recently achieved sufficient maturity for clinical applications like breast imaging. Image reconstruction is achieved with state of art iterative algorithms (full waveform inversion in the frequency domain). In this Letter, we consider a stochastic reconstruction in the time-domain. We introduce the notion of multiple super-shots and stochastic ensembles and test our inversion algorithm on publicly available experimental data. Our results show that image quality of a time-domain stochastic reconstruction may be comparable to image quality of a deterministic reconstruction in the frequency-domain, although the time-domain reconstruction is significantly slower.
	\end{abstract}
	
	\maketitle
	
	In the last two decades, USCT (ultrasound computed tomography) has drawn a lot of interest from the scientific community thanks to unprecedented image quality for breast imaging applications. Image formation is accomplished with deterministic iterative reconstruction techniques originally developed by the seismic imaging community and known as full waveform inversion in the frequency domain (FD-FWI, e.g. \cite{VirieuxOperto}); state of the art methods in the context of ring-array USCT are described in \cite{DuricLAST}. Equivalent techniques in the time-domain (TD-FWI) are indeed available, e.g.  \cite{Vetterli2}, but they are much slower with respect to their frequency-domain counterpart. For this reason, stochastic reconstructions techniques in the time-domain have been developed in the context of seismic imaging, e.g. \cite{Krebs}, and then translated to USCT in \cite{Huang}, \cite{Anastasio1},  \cite{Anastasio2}, \cite{Tromp} and \cite{Lucka}. In particular, earlier attempts to stochastic TD-FWI were considered in \cite{Anastasio1} for an experimental phantom and in \cite{Anastasio2} for patient data, but the corresponding images reconstructed with deterministic FD-FWI were not included in the analysis. In this Letter, we revisit stochastic TD-FWI by considering the concepts of multiple super-shots and stochastic ensembles recently introduced by the author in \cite{Forte} in the context of a stochastic reconstruction in the frequency domain. 	

In the following, we assume the geometry of a ring-array with $N_{_{Tx}} = N_{_{Rx}}$ transceivers. Full-waveform inversion is a model-based reconstruction technique; the physical model describes acoustics. In a lossless medium, the constant density wave equation in time-domain may be written in the continuous formulation as
	 	\begin{equation}
		\Box_{_{c}}\,p(\boldsymbol{x};\,t) = -s(\boldsymbol{x}\,;t)
		\end{equation}
		where the box operator, $\Box_{_{\boldsymbol{c}}} = \nabla^{2} - (1/c^{2})\,\partial^{2}/\partial t^{2}$, alias the Dalambertian operator, has an explicit dependence on the speed of sound $c=c(\boldsymbol{x})$. In the simplest form of TD-FWI, the only unknown model parameter is the speed of sound of the medium: the latter is updated iteratively, solving a non-linear numerical optimization problem defined by the following cost-function
				\begin{widetext}
				\begin{equation}
		C(\boldsymbol{c}) \, = \,\left\Vert \boldsymbol{d}^{^{(syn)}}(\boldsymbol{c}) - \boldsymbol{d}^{^{(obs)}}\right\Vert^{^{2}}_{_{2}}\,= \,\sum_{i_{_{Tx}} \,=\,1}^{N_{_{Tx}}}\,\left\{\sum_{j_{_{Rx}} \,=\,1}^{N_{_{Rx}}}\,\left[\sum_{k_{_{t}} \,=\,1}^{N_{_{t}}}\,\left\vert d^{^{(syn)}}_{i_{_{Tx}},\,j_{_{Rx}}}(n_{_{t}};\,\boldsymbol{c}) - d^{^{(obs)}}_{i_{_{Tx}},\,j_{_{Rx}}}(n_{_{t}})\right\vert^{^{2}}\right]\right\}
		\end{equation}
				\end{widetext}
In the previous expression, the observed data $\boldsymbol{d}^{^{(obs)}}$ are typically represented by the measured times series digitized at the sensors, a vector of size $N_{_{t}} \times N_{_{Rx}}\times N_{_{Tx}}$. By $\boldsymbol{d}_{i_{_{Tx}}}^{^{(obs)}}(\boldsymbol{t})$, we mean the $N_{_{t}} \times N_{_{Rx}}$ vector representative of the timeseries measured at all receivers when source $i_{_{Tx}}$ is active; by  $d^{^{(obs)}}_{i_{_{Tx}},\,j_{_{Rx}}}(n_{_{t}})$, we mean the $n_{_{t}}$ time sample (usually stored as a 16 bits integer) at receiver $j_{_{Rx}}$ when source $i_{_{Tx}}$ is active. The synthetic data $\boldsymbol{d}^{^{(syn)}}$, also a vector of size $N_{_{t}} \times N_{_{Rx}}\times N_{_{Tx}}$,  are obtained by solving a discretized version of the time-domain acoustic wave equation above on a computational grid of size $N_{_{x}} \times N_{_{y}}$, pixel $dx$, time step $dt$, $N_{_{t}}$ time samples and total duration $T=N_{_{t}}\,dt$; by  $d^{^{(syn)}}_{i_{_{Tx}},\,j_{_{Rx}}}(n_{_{t}};\,\boldsymbol{c})$, we mean the $n_{_{t}}$ time sample (usually stored in single or double precision) at receiver $j_{_{Rx}}$ when source $i_{_{Tx}}$ is is active; this sample is evidently dependent on the assumed speed of sound at iteration $k$, $\boldsymbol{c} = \boldsymbol{c}_{_{k}}$. In the stochastic formulation known as FWI with source-encoding, individual array elements are grouped together to form a so-called super-shot and transmit simultaneously; each array element (i.e. each source) in a given super-shot is encoded by a phase term (phase-encoding, PE). Following the terminology and the notation introduced in \cite{Forte}, we consider the general case of multiple super-shots $N_{_{ss}}$. In particular, each super-shot $ss_{_{i}}$ may contain $N_{_{Tx}}^{{ss_{_{i}}}}$ distinct elements with its own corresponding group of $N_{_{Rx}}^{{ss_{_{i}}}}$ physically distinct receiving elements, allowing to differentiate between transmission data and reflection data. The encoded cost function may be written as
					\begin{equation} \label{eq:CostEnc}
		C^{^{(enc)}}(\boldsymbol{c}) \, =\,\left\Vert \boldsymbol{D}^{^{(syn)}}(\boldsymbol{c}) - \boldsymbol{D}^{^{(obs)}}\right\Vert^{^{2}}_{_{2}}\,=
					\end{equation}
			\begin{equation*}
		 \,\sum_{ss_{_{i}} \,=\,1}^{N_{_{ss}}}\,\left\{\sum_{j_{_{Rx}} \,=\,1}^{N_{_{Rx}}^{ss_{_{i}}}}\,\left[\sum_{k_{_{t}} \,=\,1}^{N_{_{t}}}\,\left\vert D^{^{(syn)}}_{ss_{_{i}},\,j_{_{Rx}}}(n_{_{t}};\,\boldsymbol{c}) - D^{^{(obs)}}_{ss_{_{i}},\,j_{_{Rx}}}(n_{_{t}})\right\vert^{^{2}}\right]\right\}
		\end{equation*}
		with $\boldsymbol{D}^{^{(obs)}}$ and $\boldsymbol{D}^{^{(syn)}}(\boldsymbol{c})$ the encoded observed data and encoded synthetic data respectively, defined below. This approach has an enormous computational advantage, as one has to solve only $N_{_{ss}}$ wave equations to evaluate the synthetic data, and thus the cost function, compared to $N_{_{Tx}}$ wave equations in the conventional approach, but it introduces so-called cross-talk in the gradient; the cross-terms can be removed by randomly encoding the individual sources with iteration-dependent phase-encoding terms. This transforms the optimization problem into a stochastic one. As in \cite{Anastasio1}, the source-encoding may be achieved with a simple Radamacher distribution (R), e.g. $a_{i_{_{Tx}}} = \pm 1$, uniformly random; these  are generated independently per speed of sound iteration and per super-shot, the latter defined via the equation
		 	\begin{equation}
		S_{_{ss_{_{i}}}}(\boldsymbol{t}) = \sum_{i_{_{Tx}} \,=\,1}^{N_{_{Tx}}^{^{ss_{_{i}}}}}\,a_{i_{_{Tx}}}^{^{ss_{_{i}}}}\,s_{i_{_{Tx}}}(\boldsymbol{t})
		\end{equation}
Each super-shot leads to the calculation of the corresponding forward wave-field
	\begin{equation} \label{eq:forward}
		\boldsymbol{\Box}_{\boldsymbol{c}_{_{k}}}\,\boldsymbol{p}_{_{ss_{_{i}}}}(\boldsymbol{x};\,\boldsymbol{t};\boldsymbol{c}_{_{k}}) = -S_{_{ss_{_{i}}}}(\boldsymbol{t})
		\end{equation}
		In the previous expression, it is implicit that the source term is zero everywhere on the computational grid except at the interpolated grid locations of the \textit{array-elements making the super-shot}. The forward wave-field is a vector of size  $N_{_{x}} \times N_{_{y}} \times N_{_{t}}$; the restriction of the forward wave-field at the super-shot dependent receiving elements $N_{_{Rx}}^{^{ss_{_{i}}}}$ represents the encoded synthetic data, $\boldsymbol{D}^{^{(syn)}}_{_{ss_{_{i}}}}(\boldsymbol{t};\boldsymbol{c}_{_{k}})$ (vector of size $N_{_{t}} \times N_{_{Rx}}^{^{ss_{_{i}}}}$). The corresponding encoded observed data (also a vector of size $N_{_{t}} \times N_{_{Rx}}^{^{ss_{_{i}}}}$) is simply obtained by linearity through the expression
		\begin{equation}
		\boldsymbol{D}_{_{ss_{_{i}}}}^{^{(obs)}}(\boldsymbol{t}) = \sum_{i_{_{Tx}} \,=\,1}^{N_{_{Tx}}^{^{ss_{_{i}}}}}\,a_{i_{_{Tx}}}^{^{ss_{_{i}}}}\,\boldsymbol{d}_{i_{_{Tx}}}^{^{(obs)}}(\boldsymbol{t})
		\end{equation}
After solving the wave equation with encoded sources and after encoding the observed data, the synthetic data and the observed data are transformed into the (temporal) frequency domain via a DTFT (e.g. FFT), per super-shot
	\begin{equation}
		\boldsymbol{D}^{^{(syn)}}_{_{ss_{_{i}}}}(\boldsymbol{\omega};\boldsymbol{c}_{_{k}}) = \text{FFT} \left\{\boldsymbol{D}^{^{(syn)}}_{_{ss_{_{i}}}}(\boldsymbol{t};\boldsymbol{c}_{_{k}})\right\}
		\label{eq:EqPulse1}
		\end{equation}
		\begin{equation}
		\boldsymbol{D}^{^{(obs)}}_{_{ss_{_{i}}}}(\boldsymbol{\omega}) = \text{FFT} \left\{\boldsymbol{D}^{^{(obs)}}_{_{ss_{_{i}}}}(\boldsymbol{t})\right\}
		\end{equation}
		   \begin{figure*}[htbp]
    \centering 
	 \scalebox{0.53} 
	 {\includegraphics{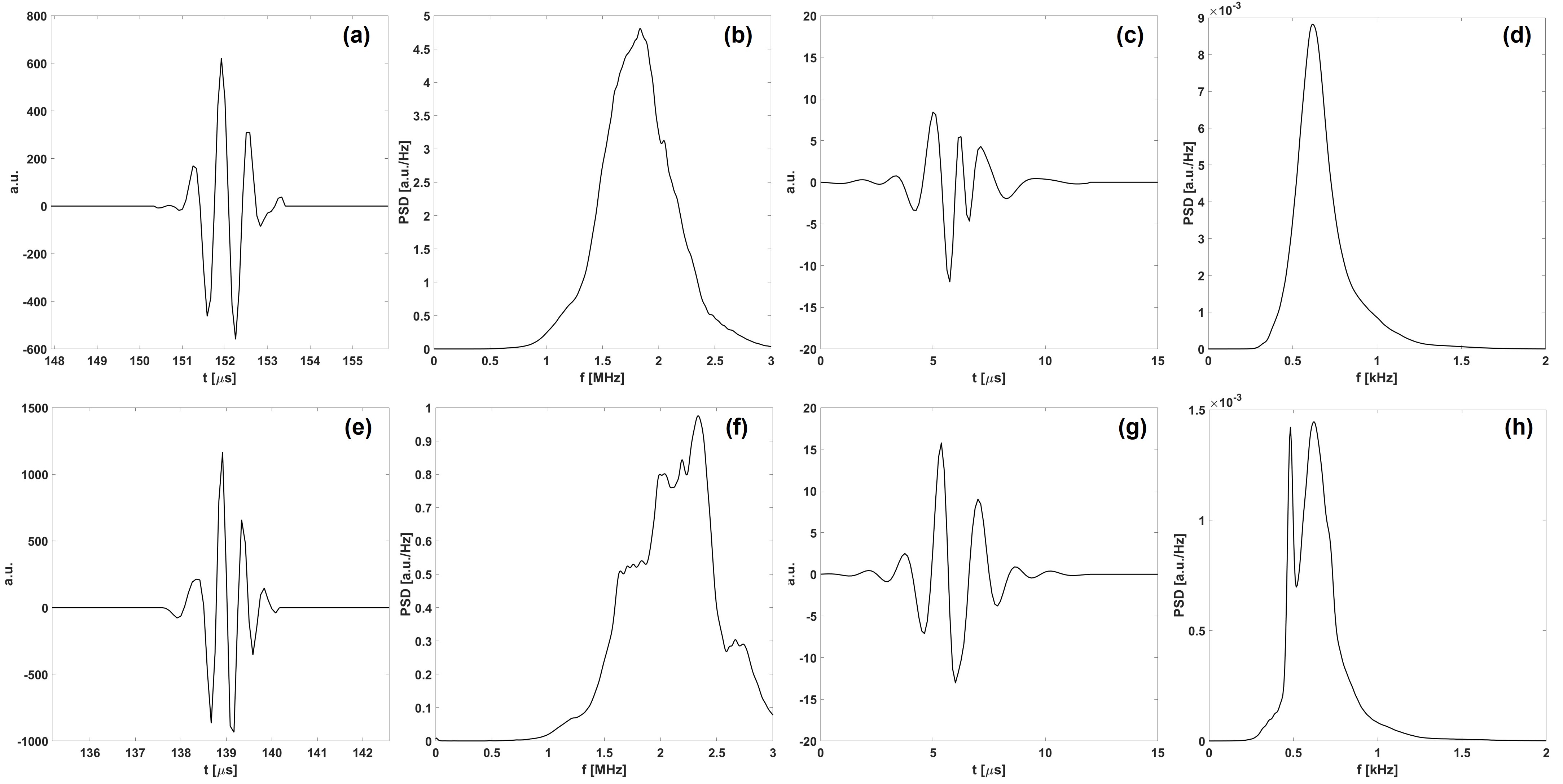}} 
	 \caption{\textit{Pulse Estimation and Spectral Contents. Raw and filtered data.} (a) Main transmission (windowed) for the malignancy raw dataset ($i_{_{Tx}} = 1\,,j_{_{Rx}} = 256$). (b) Power spectral density of transmission raw data, averaged over $j_{_{Rx}} = 128, \ldots, 384$ with $i_{_{Tx}} = 1$. (c) Time-shifted, estimated source waveform after band-pass filter. (d) Power spectral density of transmission data after band-pass filter, averaged over $j_{_{Rx}} = 128, \ldots, 384$ with $i_{_{Tx}} = 1$. Images in (e)-(h) same as (a)-(d), for the cyst dataset.}
	 \label{fig:fig1} 
	 \end{figure*}
	    \begin{figure*}[htbp]
    \centering 
	 \scalebox{0.7} 
	 {\includegraphics{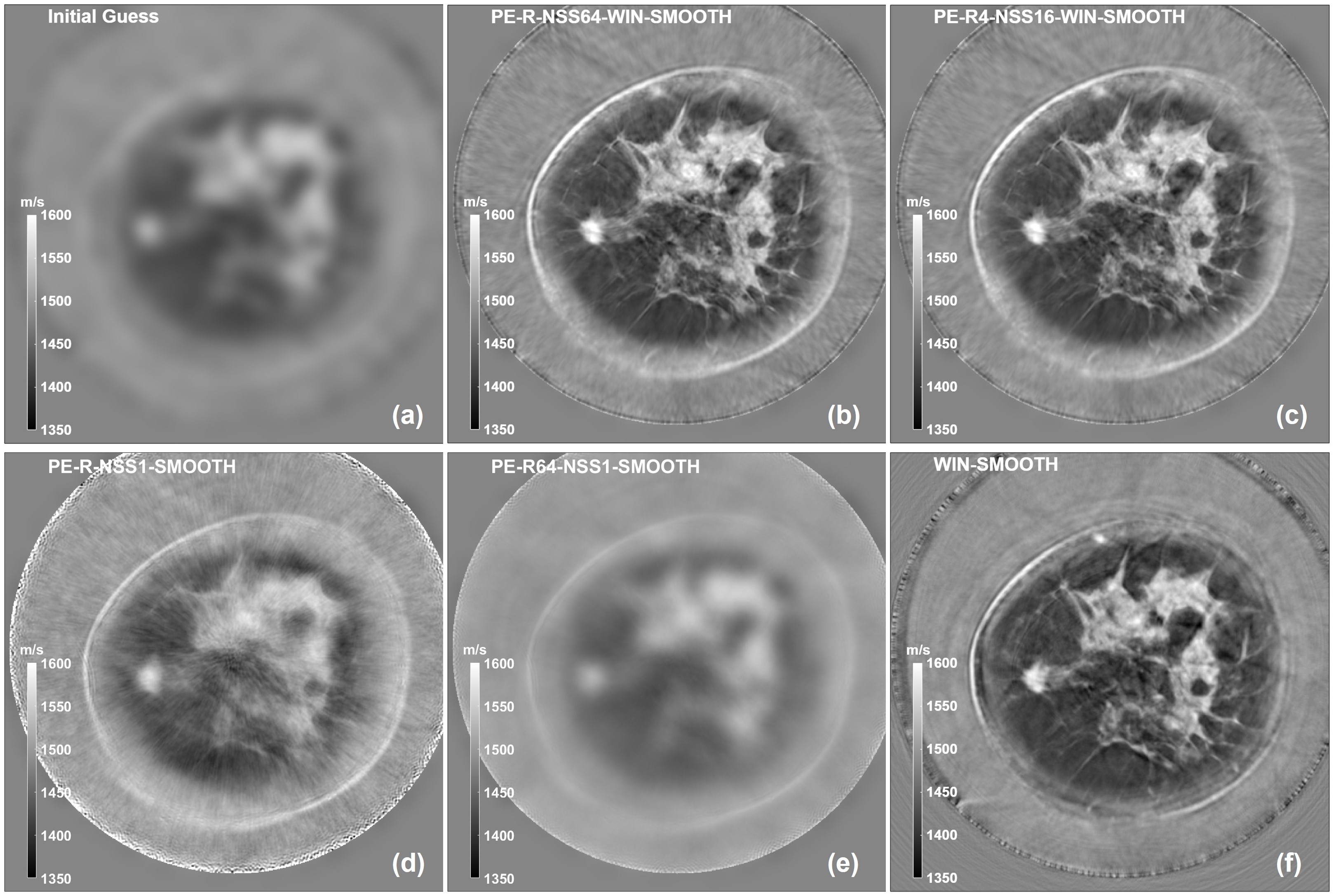}} 
    \centering 
	 \scalebox{0.7} 
	 {\includegraphics{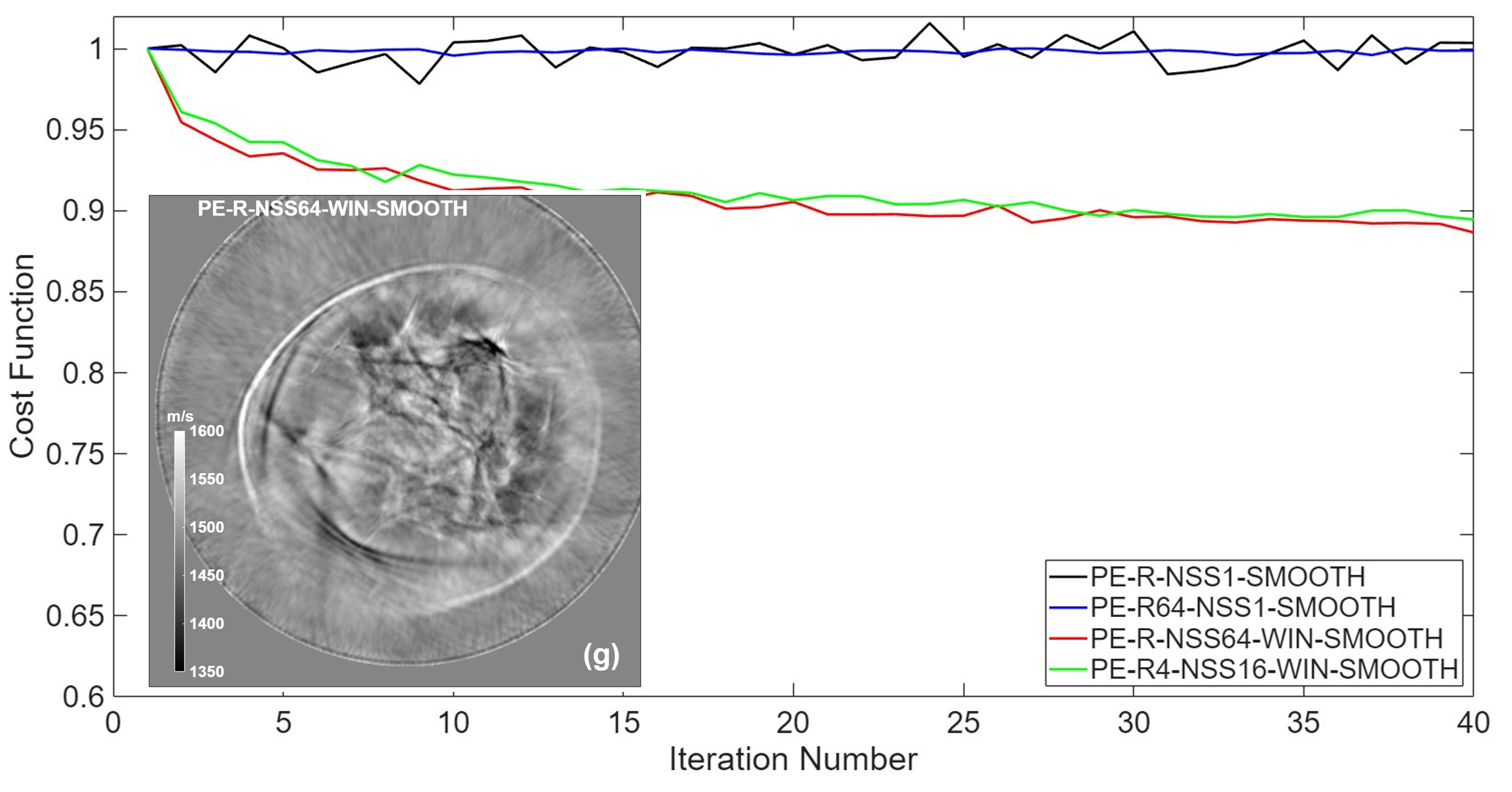}} 
	 \caption{\textit{Stochastic reconstruction with super-shots and stochastic ensembles. Reconstructed images and cost functions for experimental data (Malignancy).} (a) Initial guess for TD-reconstructions in (b)-(e). (b) TD stochastic image reconstruction with multiple super-shots ($N_{_{ss}} = 64, N_{_{PE}} = 1$). (c) TD stochastic image reconstruction with multiple super-shots ($N_{_{ss}} = 16$) and multiple stochastic ensembles ($N_{_{PE}} = 4$). (d) TD stochastic image reconstruction with one super-shot ($N_{_{ss}} = 1, N_{_{PE}} = 1$). (e) TD stochastic image reconstruction with one super-shot ($N_{_{ss}} = 1$) and multiple stochastic ensembles ($N_{_{PE}} = 64$). (f) FD deterministic image reconstruction from 300 kHz to 1 MHz from a flat initial guess. (g) TD stochastic image reconstruction with multiple super-shots ($N_{_{ss}} = 64, N_{_{PE}} = 1$) from a flat initial guess, showing cycle-skipping artifacts. TD inversions are run for 40 iterations; FD inversion is run for 10 iterations per frequency (cost function not shown). For all reconstructions, the same Gaussian smoothing filter has been applied on the final gradient before updating the speed of sound.}
	 \label{fig:M} 
	 \end{figure*}
 \begin{figure*}[htbp]
    \centering 
	 \scalebox{0.7} 
	 {\includegraphics{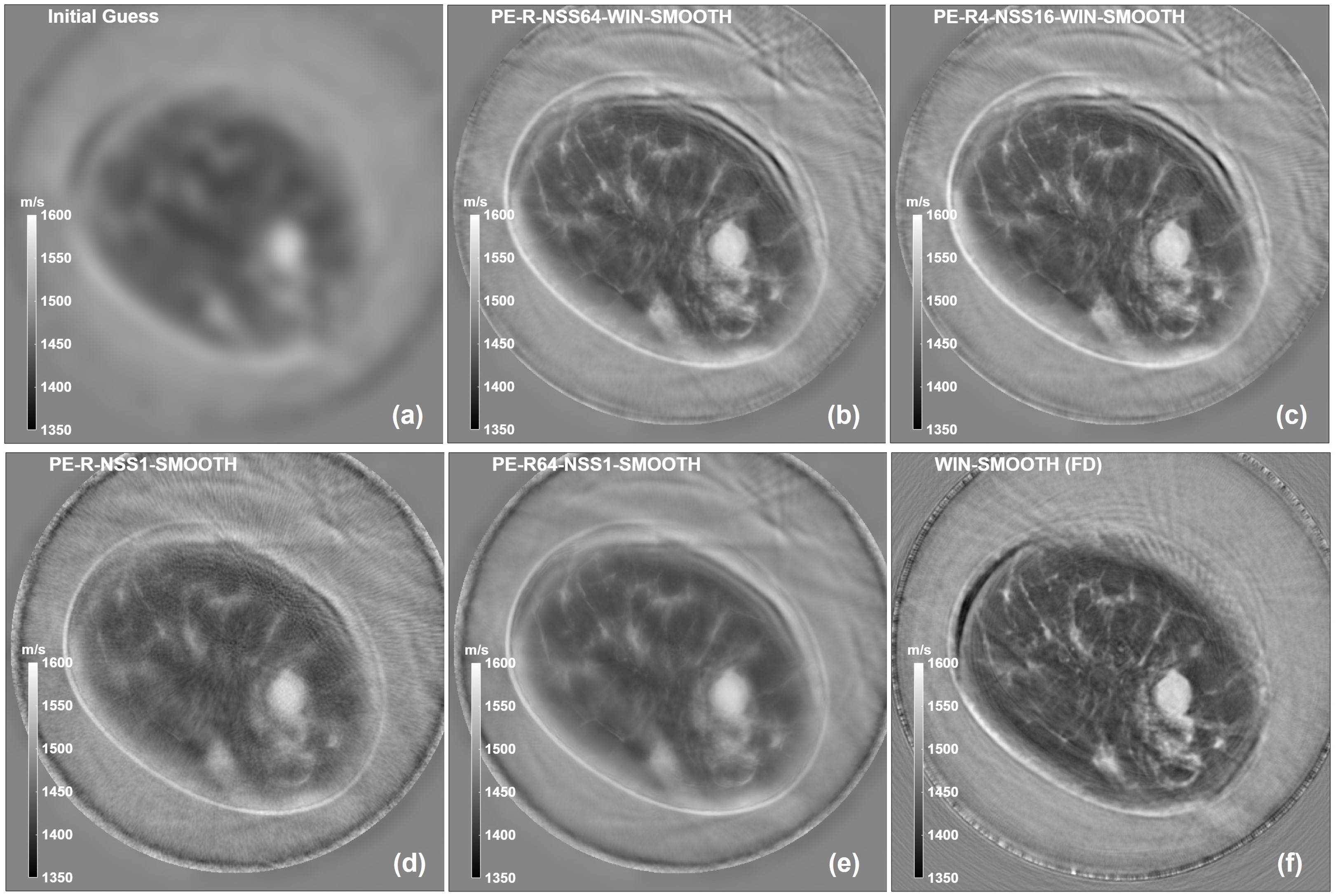}} 
    \centering 
	 \scalebox{0.7} 
	 {\includegraphics{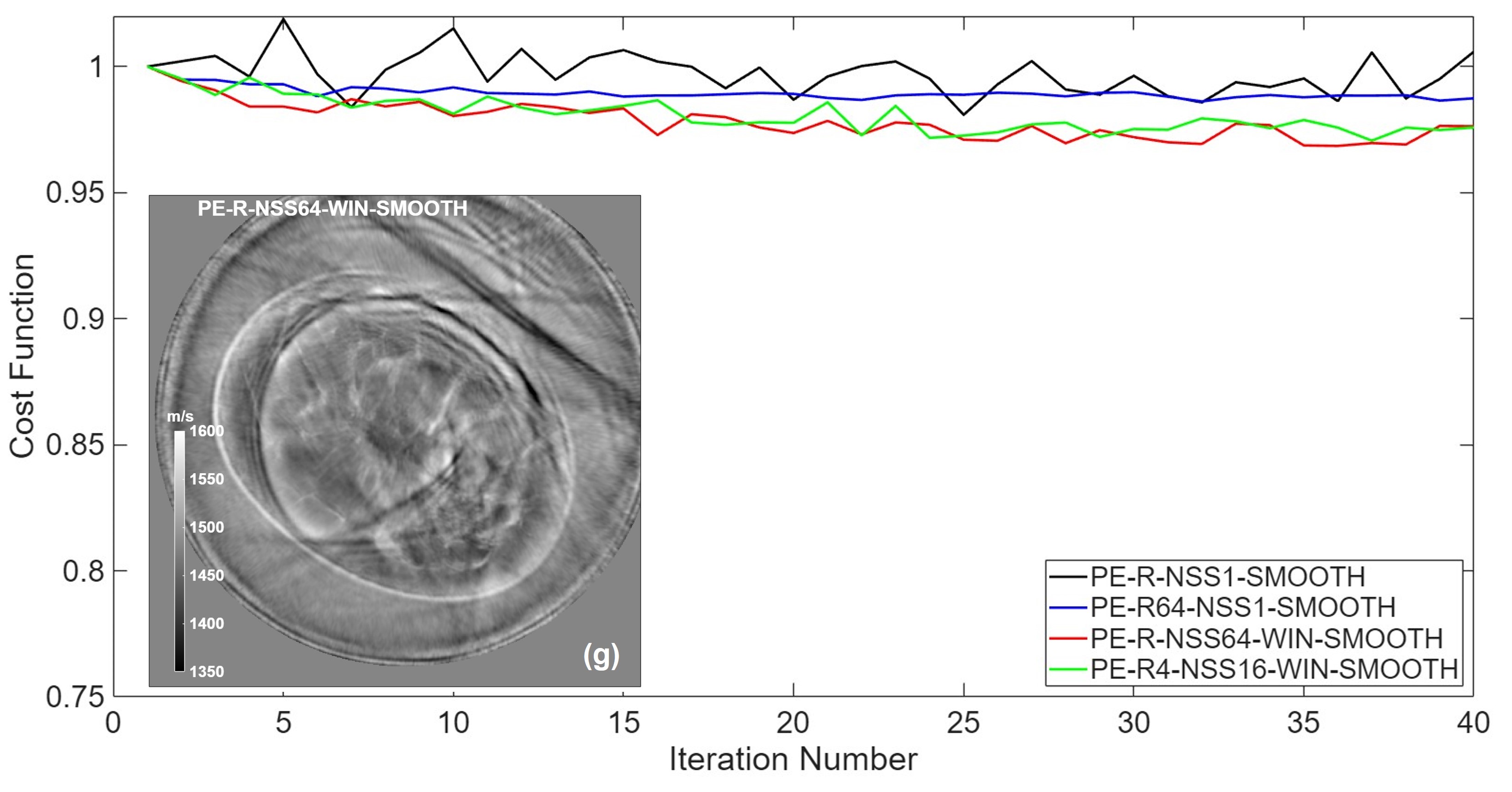}} 
	 \caption{\textit{Stochastic reconstruction with super-shots and stochastic ensembles. Reconstructed images and cost functions for experimental data (Cyst).} Same as Fig. 	\ref{fig:M}.}
	 \label{fig:C} 
	 \end{figure*}The frequency samples corresponding to the synthetic data are re-scaled to match amplitude and phase of the observed data
		\begin{equation}
		\boldsymbol{\tilde{D}}_{_{ss_{_{i}}}}^{^{(syn)}}(\omega_{_{l}};\boldsymbol{c}_{_{k}})\, =\, r_{_{ss_{_{i}}}}(\omega_{_{l}};\boldsymbol{c}_{_{k}})\, \boldsymbol{D}_{_{ss_{_{i}}}}^{^{(syn})}(\omega_{_{l}};\boldsymbol{c}_{_{k}})
		\end{equation}
		\begin{equation}
		r_{_{ss_{_{i}}}}(\omega_{_{l}};\boldsymbol{c}_{_{k}}) = \frac{\boldsymbol{D}_{_{ss_{_{i}}}}^{^{(syn)}}(\omega_{_{l}};\boldsymbol{c}_{_{k}})^{^{\boldsymbol{\star}}} \, \boldsymbol{\cdot} \boldsymbol{D}_{_{ss_{_{i}}}}^{^{(obs)}}(\omega_{_{l}};\boldsymbol{c}_{_{k}})}{\left\Vert \boldsymbol{D}_{_{ss_{_{i}}}}^{^{(syn)}}(\omega_{_{l}};\boldsymbol{c}_{_{k}}) \right\Vert^{^2}_{_{2}}} 
		\end{equation}
		where the dot product $\boldsymbol{\cdot}$ is the standard scalar product (sum along the receivers) and the $\boldsymbol{\star}$ operator is the standard complex conjugate operation.	The re-scaling factor is evaluated at all the frequency bins $\omega_{_{l}}$ that have enough SNR in the reconstruction bandwidth (see description below). The synthetic data are then re-transformed into the time-domain:
\begin{equation}
		\boldsymbol{\tilde{D}}_{_{ss_{_{i}}}}^{^{(syn)}}(\boldsymbol{t};\boldsymbol{c}_{_{k}}) = \text{IFFT} \left\{\boldsymbol{\tilde{D}}^{^{(syn)}}_{_{ss_{_{i}}}}(\boldsymbol{\omega};\boldsymbol{c}_{_{k}})\right\}
		\end{equation}
		The same process is applied to the forward wave-field:
		\begin{equation}
		\boldsymbol{p}_{_{ss_{_{i}}}}(\boldsymbol{x};\,\boldsymbol{\omega};\boldsymbol{c}_{_{k}}) = \text{FFT} \left\{\boldsymbol{p}_{_{ss_{_{i}}}}(\boldsymbol{x};\,\boldsymbol{t};\boldsymbol{c}_{_{k}})\right\}
		\end{equation}
			\begin{equation}
		\boldsymbol{\tilde{p}}_{_{ss_{_{i}}}}(\boldsymbol{x};\omega_{_{l}};\boldsymbol{c}_{_{k}})\, =\, r_{_{ss_{_{i}}}}(\omega_{_{l}};\boldsymbol{c}_{_{k}})\, \boldsymbol{p}_{_{ss_{_{i}}}}(\boldsymbol{x};\omega_{_{l}};\boldsymbol{c}_{_{k}})
		\end{equation}
			\begin{equation}
		\boldsymbol{\tilde{p}}_{_{ss_{_{i}}}}(\boldsymbol{x};\,\boldsymbol{t};\boldsymbol{c}_{_{k}})\, =\, \text{IFFT} \left\{\boldsymbol{\tilde{p}}_{_{ss_{_{i}}}}(\boldsymbol{x};\boldsymbol{\omega};\boldsymbol{c}_{_{k}})\right\}
				\label{eq:EqPulse2}
		\end{equation}
		This is indeed equivalent to solving the forward equation with a re-scaled source term, $\tilde{S}_{_{ss_{_{i}}}}(\boldsymbol{t})$.
The time-reversed residuals are computed 
	\begin{equation}
 R_{_{ss_{_{i}}}}(\boldsymbol{t};\boldsymbol{c}_{_{k}}) =  \sum_{j_{_{Rx}} \,=\,1}^{N_{_{Rx}}^{^{ss_{_{i}}}}}\, \left[ \tilde{D}_{ss_{_{i}}\, , \, j_{_{Rx}}}^{^{(syn)}}(T-\boldsymbol{t};\boldsymbol{c}_{_{k}}) -  D_{ss_{_{i}}\, , \, j_{_{Rx}}}^{^{(obs)}}(T-\boldsymbol{t}) \right]
		\end{equation}
and the adjoint equation is solved (per super-shot)
	\begin{equation} \label{eq:adjoint}
		\boldsymbol{\Box}_{\boldsymbol{c}_{_{k}}}\,\boldsymbol{q}_{_{ss_{_{i}}}}(\boldsymbol{x};\,\boldsymbol{t};\boldsymbol{c}_{_{k}})= -R_{_{ss_{_{i}}}}(\boldsymbol{t};\boldsymbol{c}_{_{k}})
		\end{equation}
			In the previous expression, it is implicit that the source term is zero everywhere on the computational grid except at the interpolated grid locations of the \textit{super-shot dependent receiving elements}. The adjoint wave-field is a vector of size  $N_{_{x}} \times N_{_{y}} \times N_{_{t}}$.
The gradient of the cost function defined in eq. (\ref{eq:CostEnc}) is the sum over all super-shots
		\begin{equation}
		\boldsymbol{\nabla_{_{\boldsymbol{c}}}}(\boldsymbol{x}) = \sum_{ss_{_{i}} \,=\,1}^{N_{_{ss}}}\, \boldsymbol{\nabla_{_{\boldsymbol{c}}}^{^{ss_{_{i}}}}}(\boldsymbol{x})
		\end{equation}
with the individual super-shot gradient given by
		\begin{equation} \label{eq:gradient}
				\boldsymbol{\nabla}_{_{\boldsymbol{c}}}^{^{ss_{_{i}}}}(\boldsymbol{x})\, = \, -\frac{2}{\boldsymbol{c}_{_{k}}^{3}}\, \sum_{n_{_{t}} \,=\,1}^{N_{_{t}}}\,\boldsymbol{q}_{_{ss_{_{i}}}}(\boldsymbol{x};\,n_{_{t}};\boldsymbol{c}_{_{k}})\,\dot{\dot{\boldsymbol{\tilde{p}}}}_{_{ss_{_{i}}}}(\boldsymbol{x};\,n_{_{t}};\boldsymbol{c}_{_{k}})\,dt
		\end{equation}
		where all the products are point-wise multiplications and $\dot{\dot{\boldsymbol{\tilde{p}}}}_{_{ss_{_{i}}}}$ is the second temporal derivative of the (re-scaled) forward wave-field. To further mitigate the cross-talk due to the individual elements making the super-shot, one may employ a strategy where multiple stochastic ensembles are generated per super-shot, for a fixed speed of sound iteration, as discussed in \cite{Forte}. In this more general case, the cost function in eq. (\ref{eq:CostEnc}) is redefined as
		\begin{widetext}
			\begin{equation}
		C^{^{(enc)}}(\boldsymbol{c}) \, =\, \sum_{ss_{_{i}} \,=\,1}^{N_{_{ss}}}\,\left\{\sum_{n_{_{PE}} \,=\,1}^{N_{_{PE}}}\left[\sum_{j_{_{Rx}} \,=\,1}^{N_{_{Rx}}^{(ss_{_{i}})}}\,\left(\sum_{k_{_{t}} \,=\,1}^{N_{_{t}}}\,\left\vert D^{^{(syn)}}_{ss_{_{i}},\,n_{_{PE}},\,j_{_{Rx}}}(n_{_{t}};\,\boldsymbol{c}) - D^{^{(obs)}}_{ss_{_{i}},\,n_{_{PE}},\,j_{_{Rx}}}(n_{_{t}})\right\vert^{^{2}}\right)\right]\right\}
		\end{equation}
				\end{widetext}
and the final gradient is simply the sum over all super-shots and over all stochastic ensembles
			\begin{equation} 
		\boldsymbol{\nabla_{_{\boldsymbol{c}}}}(\boldsymbol{x})= \sum_{ss_{_{i}} \,=\,1}^{N_{_{ss}}}\,\left\{\sum_{n_{_{PE}} \,=\,1}^{N_{_{PE}}}\, \boldsymbol{\nabla_{_{\boldsymbol{c}}}^{^{ss_{_{i}}, \,n_{_{PE}}}}}(\boldsymbol{x})\right\}
		\end{equation}
where the forward wave field and the adjoint one both have an extra dependence on the stochastic ensemble index, i.e. $ \boldsymbol{p}_{_{ss_{_{i}}, \,n_{_{PE}}}}$ and $\boldsymbol{q}_{_{ss_{_{i}}, \,n_{_{PE}}}}$, ultimately related to the encoding terms $a_{i_{_{Tx}}}^{^{ss_{_{i}}, \,n_{_{PE}}}}$ that label the stochastic ensemble $n_{_{PE}}$. For simplicity, and given the symmetry of the geometry under study, we assume the same number of stochastic ensembles for all super-shots. The speed of sound is then updated in a standard fashion following the negative direction of the gradient
\begin{equation}
\boldsymbol{c}_{_{k+1}} \, = \, \boldsymbol{c}_{_{k}} -\alpha \, \boldsymbol{\nabla_{_{\boldsymbol{c}}}}
\end{equation}
In this Letter, we have opted for a straightforward (stochastic) gradient descent algorithm with inexact line search (max 5 line searches per speed of sound iteration). 

We now present the results of the reconstruction algorithm on two experimental data-sets\cite{DuricLAST}, a patient with a malignant mass and a patient with a cyst.
The raw data show spectral contents that extend roughly from 300 kHz to 3 MHz, Fig. \ref{fig:fig1}. This is too wide for a TD reconstruction, hence the first step is to apply a band pass filter around 1 MHz. The reconstruction algorithm requires an initial estimate for the source term, $s_{i_{_{Tx}}}(\boldsymbol{t})$. This can be extracted from raw data, Fig. \ref{fig:fig1}(a) and (b); the same band-pass filter is then applied and a time-shifted, filtered pulse is finally obtained, Fig. \ref{fig:fig1}(c) and (g). The latter is replicated across all the individual elements making the super-shot. The same source term is then updated at each iteration, as described in equations (\ref{eq:EqPulse1})- (\ref{eq:EqPulse2}) above; in particular, the re-scaling factor is evaluated at all the frequency bins $\omega_{_{l}}$ in the 500 kHz - 1 MHz bandwidth, Fig. \ref{fig:fig1}(d) and (h), as these have enough SNR and are not too close to the filter cut-off frequencies. The filtering operation results into $N_{_{t}} = 1408$ time samples and a time step of $dt = 0.125\, \mu$s (8 MHz). The reconstruction grid has a size of $N_{_{x}} \times N_{_{y}}$ $(N_{_{x}} = N_{_{y}} = 875)$ and a pixel size of $dx = 0.32$ mm. The reconstruction algorithm has been implemented in MATLAB. In particular, the discretized wave equations, (\ref{eq:forward}) and (\ref{eq:adjoint}), have been solved with the popular k-Wave toolbox on the GPU (NVIDIA RTX A4000 16 GB); the rest of the algorithm runs on the CPU in single precision. The forward and adjoint wave-fields, $\boldsymbol{p}$ and $\boldsymbol{q}$ respectively, are natively vectors of size $N_{_{x}} \times N_{_{y}} \times N_{_{t}}$; as suggested in \cite{Anastasio1}, to reduce memory requirements due to the temporal dimension of these two vectors, both are stored only on a partial FOV (field of view), the latter essentially coinciding with the region bounded by the ring-array. This allows to evaluate the gradient, eq. (\ref{eq:gradient}), only on the chosen FOV and ultimately updating the speed of sound only on this FOV. To cope with experimental noise, a mild Gaussian smoothing filter (SMOOTH) has been applied on the final gradient before updating the speed of sound. When employing multiple super-shots, the receivers are selected by applying a rectangular window (WIN) across all possible array elements. A direct implementation of the reconstruction algorithm with multiple super-shots ($N_{_{ss}} = 64$) starting from a  flat initial guess shows the known phenomenon of cycle-skipping, Fig. \ref{fig:M}(g) and Fig. \ref{fig:C}(g). Cycle-skipping is a form of non-linear aliasing that creates ghost structures vaguely resembling the true ones; this is due to the simultaneous inversion of a large frequency bandwidth, with the higher frequencies responsible for this behavior. To overcome this limitation, the inversion has to start from a non-trivial initial velocity model; the standard approach is to create a blurred image running a travel-time tomography reconstruction algorithm followed by standard FWI. Here we adopt a more practical approach where the initial guess is obtained by running a standard FD-FWI reconstruction algorithm up to 400 kHz\cite{DuricLAST}; the resulting image is severely blurred before starting the TD-FWI reconstruction, Fig. \ref{fig:M}(a) and Fig. \ref{fig:C}(a). The reconstructed images obtained with a single super-shot ($N_{_{ss}} = 1, N_{_{Tx}}^{{ss}} =  N_{_{Rx}}^{{ss}} =  N_{_{Tx}} = N_{_{Rx}} = 512 , N_{_{PE}} = 1$), Fig. \ref{fig:M}(d) and Fig. \ref{fig:C}(d), partially show the true distribution of the speed of sound of the breast slice. \textit{However}, both show a peculiar image texture in the form of radial stripes (spokes), as already observed in \cite{Anastasio1}-\cite{Anastasio2}; this texture doesn't disappear with increasing the number of iterations, it actually becomes more prominent. The main source of this texture is the residual cross-talk due to phase-encoding. To test this hypothesis, the reconstruction is run again with one super-shot ($N_{_{ss}} = 1, N_{_{Tx}}^{{ss}} =  N_{_{Rx}}^{{ss}} =  N_{_{Tx}} = N_{_{Rx}} = 512 $) and multiple stochastic ensembles ($N_{_{PE}} = 64$), Fig. \ref{fig:M}(e) and Fig. \ref{fig:C}(e): this is indeed effective in removing any image texture, as anticipated in the Conclusions section in \cite{Forte}. However, whereas the cyst dataset does show improved image quality, the malignancy one shows very little improvement. This is an indication that a single super-shot may not be reliable and that results may be data-dependent. We believe that the main reason behind this is that with a single super-shot it's not possible to differentiate between reflection data and transmission data. The reconstruction is then run again, with multiple super-shots ($N_{_{ss}} = 64, N_{_{Tx}}^{{ss_{_{i}}}} = 86, N_{_{Rx}}^{{ss_{_{i}}}} = 384, N_{_{PE}} = 1$), Fig. \ref{fig:M}(b) and Fig. \ref{fig:C}(b), and in combination with stochastic ensembles ($N_{_{ss}} = 16, N_{_{Tx}}^{{ss_{_{i}}}} = 86, N_{_{Rx}}^{{ss_{_{i}}}} = 384, N_{_{PE}} = 4$), Fig. \ref{fig:M}(c) and Fig. \ref{fig:C}(c). For comparison, images reconstructed on the same computational grid with a deterministic inversion in the frequency domain are also shown (10 iterations per frequency, from 300 kHz to 1 MHz in steps of 100 kHz), Fig. \ref{fig:M}(f) and Fig. \ref{fig:C}(f). These results are in perfect agreement with what we have shown in \cite{Forte} for the stochastic inversion in the frequency-domain.
With all the caveats and the difficulties in assessing image quality, we feel that it is fair to claim that both TD and FD reconstructions display the same anatomy; in particular the \textit{relative} contrast among the different tissues is preserved in all cases, although the TD reconstruction looks soft. There is, \textit{however}, a significant difference in terms of computing time and resources. Whereas a FD deterministic inversion may take 60-90 minutes on a single GPU\cite{DuricLAST} and a FD stochastic one is much faster and provides comparable image quality \cite{Forte}, a TD stochastic reconstruction may take more than 24 hours with the same hardware architecture (by keeping fixed the product the product $ N_{_{ss}} \times N_{_{PE}} = 64$), from a non-trivial initial guess; this explains why the TD stochastic reconstruction has been run only for 40 iterations. These considerations may rule out the feasibility of a stochastic inversion in time-domain in ring-array USCT, let alone any deterministic inversion in time-domain.

Earlier attempts to TD randomized inversions for the geometry of a ring-array have been considered in \cite{Anastasio1} for an experimental phantom and in \cite{Anastasio2} for patient data (the latter \textit{seems} to be the same cyst slice considered in this Letter). In these two papers, the authors have considered only one super-shot, approach that doesn't allow to distinguish between transmission data and reflection data. In an attempt to mitigate the impact of reflection data, forward data were contaminated with synthetic data; the same pulse estimated from experimental water-bath data was employed for all transmitting elements and for all iterations, unlike the approach described in equations (\ref{eq:EqPulse1})- (\ref{eq:EqPulse2}) above that allows to adapt the spectral contents of the pulse to the sensors data (per super-shot and per-iteration). We believe that the difference in image quality between these two papers and the current Letter can indeed be explained by their approach relying on a single super-shot and a more naive estimation for the source waveform.

A fair comparison between TD- and FD-FWI is far from being trivial. In fact, in any TD inversion all frequencies above noise contribute to image formation, whereas in a FD inversion only a discrete number of pre-selected frequencies contribute to image formation. In the present case, the spectra of the signals show a non-zero SNR up to 1.5 MHz, Fig. \ref{fig:fig1}(d) and (h) , whereas in the FD inversion the maximum inverted frequency has been set to 1 MHz. On the other hand, the filtering operation may decrease the SNR at the frequencies employed for the TD inversion, whereas no operation is needed on the raw data except a DTFT for the FD inversion, thus preserving the native SNR. The filtering operation itself introduces extra complications, as the frequencies close to the filter cut-off frequencies may be distorted with potentially detrimental impact on image quality. For these reasons, it is generally agreed that TD inversions exhibit additional difficulties and require much longer computing times to achieve comparable image quality. Nevertheless, the effectiveness of a time-domain randomized inversion with source-encoding on patients' data and a detailed comparison with the frequency-domain inversion have been shown for the first time in this Letter. 
	\bibliography{TDReconSupershotsAX}
	
	\end{document}